\documentclass[12pt]{iopart}
\pdfoutput=1
\usepackage{amsfonts}
\usepackage{graphicx}
\usepackage{iopams}
\usepackage[mathscr]{euscript}
\usepackage{ragged2e} 

\usepackage{xcolor}
\usepackage{bm}
\usepackage{subfigure}
\usepackage{textcomp}
\usepackage{amsmath}
\usepackage{cite}
\usepackage{float}
\usepackage{soul}
\usepackage{stackrel}
\usepackage{ulem}
\usepackage{multirow}
\usepackage{MnSymbol}
\usepackage{cancel}

\usepackage{stackengine}

\usepackage{booktabs}

\usepackage{tikz}
\usepackage{esint}
\usetikzlibrary{arrows,arrows.meta}
\newcommand\semiInt[1][1]{%
    \begin{tikzpicture}[scale=0.2]
        \coordinate (center) at (0.2,0.55);
        \draw[black,-{>[scale=0.6]}] (0, 0.0) + (center) arc (180:0:0.5);
       $\int$
    \end{tikzpicture}
    }

\usepackage[colorlinks,linkcolor=blue,urlcolor=blue,citecolor=blue]{hyperref}
\allowdisplaybreaks

\pdfoutput=1
\usepackage{esint}

\usepackage{hyperref}

\begin{document}

\title{{\color{black}Non-stationary Statistics and} Energetics of Brownian Motion under Stochastic Harmonic Confinement}

\author{Deepak Gupta$^{1,2}$ and Sabine H. L. Klapp$^{2}$}
\address{$^{1}$Institute of Physics, Polish Academy of Sciences, Al. Lotnik\'ow 32/46, 02-668,
Warsaw, Poland}
\address{$^{2}$Institut f\"ur Physik und Astronomie, Technische Universit\"at Berlin, Hardenbergstrasse 36, D-10623 Berlin, Germany}
\ead{phydeepak.gupta@gmail.com}

\vspace{10pt}

\begin{abstract} 
We investigate the positional statistics and thermodynamic properties of a Brownian particle confined in a Harmonic trap. The stiffness of the particle fluctuates in time as the square of the Brownian process. 
First, in the absence of a thermal bath, we investigate the probability density function of the position of the particle at time $t$. Then, we provide an expression to compute the $n$th positional moment. We evaluate the first four positional moments and discuss their asymptotic behavior in the long-time limit. {\color{black}In contrast to previously studied models of fluctuating stiffness, the current model describes a non-stationary process.} Furthermore, we provide the exact expression for the average work performed on the system and the average heat exchanged by the particle with the bath. {\color{black}Their long time behavior again reflects the non-stationary nature of the process.} Our theoretical predictions are supported by numerical simulations. 
\end{abstract}

%
\vspace{2pc}
\noindent{\it Keywords}: 

\section{Introduction}
\label{intro} 
The Langevin equation~\cite{Langevin_Paul} is one of the fundamental equations for describing stochastic dynamics and has found applications across a broad range of disciplines, including mathematics~\cite{math}, finance~\cite{Bouchaud1998}, chemistry~\cite{KRAMERS1940284, Chemical-LE}, physics~\cite{sekimoto2010stochastic,Spinney-Ford}, and biology~\cite{Seifert_2012-ST}. This equation was originally introduced to describe the erratic motion of microscopic particles suspended in a fluid~\cite{Langevin_Paul}; it has since become a universal tool for modeling systems subjected to thermal fluctuations. Beyond characterizing the dynamics itself, the Langevin equation serves as a starting point for investigating the system's thermodynamic properties, such as the work performed on the system, the heat exchanged by the system with the environment, and entropy production~\cite{sekimoto2010stochastic,Seifert_2012-ST,Spinney-Ford, Jarzynski_1997,HF-2}.

Over the past two decades, advances in experimental techniques have enabled the direct observation of stochastic trajectories of microscopic systems~\cite{ciliberto_exp}, such as colloidal particles~\cite{colloidal_particle,ciliberto_exp}, molecular motors~\cite{mishima,mm-2}, and electronic circuits~\cite{EC-1,EC-2}. These developments have provided extensive verification of Langevin dynamics, including the estimation of thermodynamic observables within the framework of stochastic thermodynamics~\cite{ciliberto_exp, EC-1, mishima,colloidal_particle,Exp-6}.


In recent years, researchers have extended the applicability of the Langevin equation
to a special class of stochastic systems experiencing disordered and fluctuating environments. In such situations, the parameters entering the Langevin equation are themselves random variables or stochastic processes. Broadly speaking, two classes of disorder, not only limited to Langevin systems, have been extensively studied: {\it quenched} and {\it annealed} disorder.

In systems with quenched disorder, the random parameters remain fixed during the observation time but vary across different realizations of the system. Consequently, one considers an ensemble of systems, each characterized by a particular realization of the disorder, and physical observables are obtained by averaging over the disorder ensemble. Examples include diffusion in random energy landscapes~\cite{Jan-M-1}, spin glasses~\cite{quench-2}, ecological communities~\cite{quench-4,population-quench,galla2024generatingfunctionalanalysisrandomlotkavolterra}, the flocking of active particles~\cite{quench-3}, phase transitions in the random-field Ising model~\cite{qunech-1}, and pattern-forming systems~\cite{Yizhaq_2016}.

Annealed disorder, on the other hand, corresponds to situations in which the disorder evolves dynamically on time scales comparable to those of the system itself. In this case, the fluctuating environment continuously interacts with the system, leading to richer dynamical and thermodynamic behaviors. Since the disorder changes with time, physical observables must be averaged over both thermal noise trajectories and disorder realizations. Examples of such systems include diffusion with fluctuating diffusivity~\cite{DF-1,DF-2,DF-3,Jain2017,SNGSdiffusion,Santra_2022, MK-SK}, particles moving in fluids with stochastic viscosity~\cite{ROZENFELD1998409,Huang2020},  systems with fluctuating masses~\cite{gitterman2012oscillator-1,random_mass_1,Huang2020}, colloidal or quantum particles trapped in stochastically modulated confining potentials~\cite{Alston_2022,Gomez-Solano_2010, Apal-w-1,HaenggiBartussek1996, DOERING19981, Shuttling-1,Shuttling-2}, finite-time stochastic resetting protocols~\cite{reset_linear,SRSR-pal,goerlich2024tamingmaxwellsdemonexperimental,Besga_2020} to analyze their associated thermodynamic costs~\cite{olsen2024thermodynamic,gupta2022work, olsen2024thermodynamic2, gupta2025thermodynamiccostrecurrenterasure}, and consumer-resource~\cite{zanchetta2025emergenceecologicalstructurespecies} and Lotka-Volterra models~\cite{LVA-1}.

Motivated by recent advances in the study of annealed disorder, we recently investigated a model in which a Brownian particle is confined in a harmonic trap whose stiffness fluctuates according to a stationary Ornstein–Uhlenbeck process~\cite{Gupta_2025-SHT} (see also Ref.~\cite{Cocconi_2024} for a similar analysis). In that work, exact analytical expressions were obtained for the positional moments of the particle at arbitrary times. Furthermore, within the framework of stochastic thermodynamics, we calculated the average work performed on the particle by the fluctuating trap and characterized the associated energetic exchanges.
However, a limitation of the Ornstein-Uhlenbeck description is that the stiffness fluctuations are Gaussian and therefore allow both positive and negative values. While mathematically convenient, negative stiffness values correspond to transiently unstable trapping potentials and may not be physically relevant for many experimental systems. In realistic situations, the stiffness of a confining potential is typically non-negative and may fluctuate due to changes in environmental conditions, external driving, or intrinsic fluctuations of the trapping mechanism.

To address this issue, we consider a Brownian particle confined in a harmonic trap, whose stiffness is modeled by the square of a Brownian motion. Notice that here stiffness is a non-stationary process and remains strictly non-negative while retaining the stochastic nature of the confinement. The resulting dynamics provide a physically motivated example of a Langevin system subjected to multiplicative annealed disorder.

Our 
goal is to characterize both the dynamical and thermodynamic properties of this system. 
In this work, the stiffness is a stochastic non-Gaussian process; therefore, the analytical treatment of this model differs from that of the previous models~\cite{Cocconi_2024, Gupta_2025-SHT}. Here, we employ the Feynman-Kac formalism to obtain exact analytical expressions for the first four positional moments (quantifying the influence of the fluctuating confinement) of the particle for arbitrary times~\cite{kac1949distributions, SNM-Brownian-functionals,perez2012feynman}.
We emphasize that, unlike previously studied models of fluctuating stiffness~\cite{Cocconi_2024, Gupta_2025-SHT} where the fluctuations of stiffness approach a stationary distribution in the long-time limit, the present model exhibits {\it non-stationary positional fluctuations.} In addition, using the framework of stochastic thermodynamics, we analytically calculate the average work performed on the particle by the stochastic modulation of the trap stiffness and the average heat exchanged between the particle and the surrounding thermal reservoir. These results provide new insights into energy transfer processes in stochastic systems driven by positive-valued fluctuating stiffness and contribute to the broader understanding of nonequilibrium thermodynamics in disordered systems.

\section{Model}
\label{sec:model}
We consider a Brownian particle in a harmonic trap. The stiffness of the trap has two contributions: 1) Static stiffness, $\sigma_0 \kappa$, characterized by a time-independent parameter $\sigma_0$; 2) stochastic stiffness, $\sigma k(t)$, where the strength of this stochasticity is captured by the dimensionless parameter $\sigma$. From the physical point of view, the fluctuating stiffness should not become negative. For simplicity, we here consider $k(t)=a^2(t)$, where $a(t)$ is a Brownian process. Therefore, the following coupled Langevin dynamics describe the system
\begin{subequations}
\label{main-xk-eqn}
\begin{align}
    \dot x(t) &= - [\sigma_0 t_k^{-1}  + \sigma \gamma^{-1}k(t)] x(t) + \sqrt{2D}\eta(t) \ , \label{x-eqn}\\
    k(t)&=\kappa a^2(t)\ , \\
     \dot a(t) &=  \sqrt{t_k^{-1}}\xi(t)\label{a-eqn} \ ,
\end{align}
\end{subequations}
where the dot denotes a time derivative, and  $t_k\equiv \gamma/\kappa$ is the relaxation time of the harmonic trap in the absence of stochastic stiffness $k(t)$ with a friction constant $\gamma$. In addition, $\eta(t)$ and $\xi(t)$ are Gaussian white noises with zero mean and delta correlation in time, i.e., $\langle \eta(t)\eta(t')\rangle=\langle \xi(t)\xi(t')\rangle =\delta(t-t')$. Moreover, the noises $\eta(t)$ and $\xi(t)$ are independent of each other. Here, the brackets $\langle \cdots \rangle$ indicate the ensemble average over the noise realizations. In Eq.~\eqref{x-eqn}, $D=k_{\rm B}T/\gamma$ is the diffusion constant, where $T$ is the temperature of the environment, and $k_{\rm B}$ is Boltzmann's constant.

We rescale the time $t/t_k\to t$ and the stiffness constant $k(t)/\kappa\to k(t)$, yielding
\begin{subequations}
\label{main-xk-eqn-2}
\begin{align}
    \dot x(t) &= - [\sigma_0 + \sigma k(t)] x(t) + \sqrt{2D t_k}\eta(t) \ , \label{x-eqn-2}\\
    k(t) &= a^2(t)\ , \\
     \dot a(t) &= \xi(t)\label{a-eqn-2} \ .
\end{align}
\end{subequations}
The above stochastic differential equations are supplemented by initial conditions $x(0)=x_0$ and $a(0)=a_0$. For simplicity, we set $a_0=0.$

Before we proceed to calculate the fluctuations of $x(t)$, we first compute the probability density function of $k(t)$. Since the process $\xi(t)$~\eqref{a-eqn-2} is Gaussian, the probability density function of $a(t)$ is Gaussian
\begin{align}
    p(a,t|a_0=0) = \dfrac{e^{-\frac{a^2}{2t}}}{\sqrt{2 \pi t}}\ .
\end{align}
Therefore, the probability density function of $k(t)$ can be straightforwardly written as
\begin{align}
    p(k,t) = \int_{-\infty}^{\infty}~da\dfrac{e^{-\frac{a^2}{2t}}}{\sqrt{2 \pi t}}~\delta(k-a^2)\ .
\end{align}
\begin{figure}
    \centering
    \includegraphics[width=0.5\linewidth]{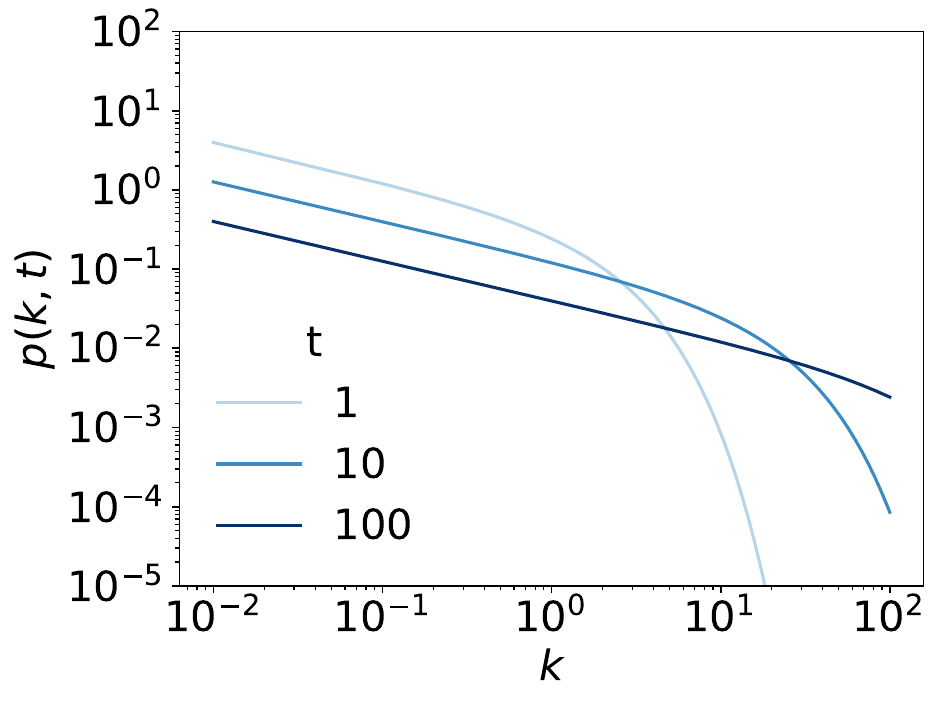}
    \caption{Probability density function $p(k,t)$~\eqref{eq:pkt} as a function of stiffness for different times $t$.}
    \label{fig:pkt}
\end{figure}
We proceed by substituting the following expression for the delta function on the right-hand side of the above equation
\begin{align}
    \delta(f(a)) = \sum_{a_i} \frac{\delta(a - a_i)}{\left| f'(a_i) \right|} = \dfrac{1}{2\sqrt{k}} [\delta(a - \sqrt{k}) + \delta(a + \sqrt{k})]\ ,
\end{align}
from which we obtain
\begin{align}
    p(k,t) = \dfrac{e^{-\frac{k}{2t}}}{\sqrt{2\pi k t}}\ .\label{eq:pkt}
\end{align}
The above distribution is a Gamma distribution $p(k;\, \alpha,\theta)=\frac{k^{\alpha - 1} e^{-k/\theta}}{\Gamma(\alpha)\,\theta^{\alpha}}$ with shape and scale parameters, respectively, $\alpha = 1/2$ and $\theta =2t$ (see Fig.~\ref{fig:pkt}). Furthermore, given the evolution equation for $a(t)$~\eqref{a-eqn-2}, we can compute the mean and correlation of $k(t)$. These are, respectively, given by
\begin{subequations}
\label{k-mom}
\begin{align}
    \langle k(t)\rangle  &= \langle a^2(t)\rangle = \int_0^t~ds_1\int_0^t~ds_2~\langle \xi(s_1)\xi(s_2)\rangle = t\ ,\label{eq:k-avg} \\
   \langle \delta k(t_1)\delta k(t_2)\rangle  &=2\begin{cases}
        t^2_2 \qquad t_1>t_2\\
         t_1^2\qquad t_2>t_1\\
    \end{cases}\ .
\end{align}    
\end{subequations}

Similarly, we can compute the higher order correlations. As we show in the following sections, even if we know the distribution of $k(t)$ and its correlations, it is not straightforward to compute the distribution of $x(t)$, as it requires averaging over the trajectories of noise $\eta(t)$ and $k(t)$. In section~\ref{sec:T=0}, we specialize our calculations for the case when there is no thermal bath, i.e., $T=0$. Then, in Sec.~\ref{sec:gen_tp}, we will discuss the case when $T\neq 0$.

\section{Computation of Moment Generating Function}
\label{sec:mgf}
Before we discuss the positional fluctuations for $T=0$~(Sec.~\ref{sec:T=0}) and $T\neq 0$~(Sec.~\ref{sec:gen_tp}), we first compute a relevant quantity that will be helpful for the subsequent calculations.
Specifically, our aim is to compute the following conditional moment generating function 
\begin{align}
    Q(a_s,s) = \langle e^{-\int_{s}^t \, dr~q(r)a^2(r)}\rangle\ , \label{cond-avg}
\end{align}
where the angular brackets indicate the average over trajectories starting from $a_s$ at time $s$ to $a$ at time $t$, with the evolution of $a(t)$ given by Eq.~\eqref{a-eqn-2} ($a(t)$ is a Brownian process). In the above equation~\eqref{cond-avg}, $q(r)$ is a time-dependent parameter (later in Secs.~\ref{sec:T=0} and \ref{sec:gen_tp}, we will discuss its specialized forms). From Eq.~\eqref{cond-avg}, we find the terminal condition $Q(a_s,t)=1$. The full moment generating function is then obtained by averaging over trajectories starting from $a=0$, i.e., 
\begin{align}
    Q_{\rm full}(s) = \int_{-\infty}^\infty~da~p(a_s,s|a=0,t=0)~Q(a_s,s)\ ,\label{Q-full}
\end{align}
where $p(a_s,s|0,0)$ is the propagator corresponding to trajectories starting from $a=0$ at time $t=0$ to $a_s$ at time $t=s$. In our case, $p(a_s,s|0,0)$ is the propagator of the diffusion equation corresponding to Eq.~\eqref{a-eqn-2}. Thus, the above equation can be rewritten as 
\begin{align}
 Q_{\rm full}(s) = \int_{-\infty}^\infty~da~\dfrac{e^{-\frac{a_s^2}{2 s}}}{\sqrt{2 \pi s}}~Q(a_s,s)\ .   \label{fin-avg} 
\end{align}
This full moment generating function $Q_{\rm full}(s)$ will be used to compute the positional fluctuations in Secs.~\ref{sec:T=0} and \ref{sec:gen_tp}.

Corresponding to Eq.~\eqref{cond-avg} and the evolution equation~\eqref{a-eqn-2}, we write the backward Fokker-Planck equation, i.e., the Feynman-Kac formula~\cite{kac1949distributions, SNM-Brownian-functionals,perez2012feynman},
\begin{align}
    -\partial_u Q(a_s,u) = \frac{1}{2}\partial_{a_s}^2 Q(a_s,u) - q(u) a_s^2 Q(a_s,u)\ ,\label{FKE}
\end{align}
for $s\leq u\leq t$. 
To solve the above equation~\eqref{FKE}, we substitute the following Gaussian ansatz
\begin{align}
    Q(a_s,u) = \exp[-\alpha(u)a_s^2 + \beta(u)]\ ,  \label{Gau-ans}
\end{align} on both sides. This yields 
\begin{align}
    \dot \alpha(u) a_s^2 -\dot \beta(u) = -\alpha(u) +2\alpha^2(u) a_s^2 - q(u) a_s^2 \ ,
\end{align}
where the dot denotes the derivative with respect to $u$. By comparing the coefficients of different powers of $a_s$, we obtain two first order differential equations for the functions $\alpha(u)$ and $\beta(u)$. These are given by
\begin{subequations}
\label{alp-bet}
    \begin{align}
    \dot \alpha(u) &=  2\alpha^2(u) -q(u)\ ,\label{alph-eqn}\\
    \dot \beta(u) &= \alpha(u)\ ,\label{beta-dff}
\end{align}
\end{subequations}
where Eq.~\eqref{alph-eqn} is the Riccati equation~\cite{ince2012ordinary}. The differential equations~\eqref{alp-bet} are supplemented by the terminal conditions $\alpha(t)=0$ and $\beta(t)=0$ [Eq.~\eqref{Gau-ans}], since $Q(a_s,t)=1$. Further, substituting
\begin{align}
    \alpha(u) = -\dfrac{1}{2}\dfrac{\dot y(u)}{y(u)} \label{sub-alp}
\end{align} 
translates the non-linear Riccati equation~\eqref{alph-eqn} to a second order linear differential equation~\cite{ince2012ordinary}
\begin{align}
    \ddot y(u) - 2q(u) y(u)=0\ . \label{y-diff}
\end{align}
To solve the above equation backwards, we require the terminal conditions for $y(u)$ and its derivative $y'(u)$ at $u=t$. Given the terminal condition $\alpha(t)=0$, we immediately have $\dot y(t)=0$ [see Eq.~\eqref{sub-alp}]. In addition, $\beta(t)=0$ gives $y(t)=1$. 

Now, for a given $q(u)$ in Eq.~\eqref{cond-avg}, we can solve the differential equations~\eqref{y-diff}. Substituting this solution into Eq.~\eqref{sub-alp} gives $\alpha(u)$, and then, using Eq.~\eqref{beta-dff}, we obtain $\beta(u)$. Furthermore, substituting $\alpha(u)$ and $\beta(u)$ into Eq.~\eqref{Gau-ans} gives us $Q(a_s,u)$. Then, substituting $Q(a_s,s)$ into Eq.~\eqref{fin-avg} gives $Q_{\rm full}(s)$.

\section{Temperature $T=0$}
\label{sec:T=0}
In this section, we focus on the case where the trapped particle is not coupled to a heat bath. However, there is still stochastic behavior arising from the fluctuations of the trap stiffness. Then, the equations of motion~\eqref{main-xk-eqn-2} reduce to
\begin{subequations}
\label{xk-eqn-3}
    \begin{align}
    \dot x &= - [\sigma_0  + \sigma a^2(t)] x(t) \ ,\label{T0xeqn}\\ 
     \dot a &=  \xi(t)\ . \label{a-eqn18}
\end{align}
\end{subequations}

To obtain the corresponding distribution of $x(t)$, we write $X(t) = \ln x(t)$, and then solve Eq.~\eqref{T0xeqn}, which gives the following
\begin{align}
    X(t) &= X_0 - \sigma_0 t - \sigma \int_0^t~ds~k(s) = X_0 - \sigma_0 t - \sigma\underbrace{\int_0^t~du~a^2(u)}_{Z(t)}\ , \label{y-eqn}
\end{align}
where the right-most term, $Z(t)$, is a functional of the trajectories of $a(t)$~\eqref{a-eqn18}. Thus, to compute the distribution of $X(t)$, we need to evaluate the distribution of $Z(t)$.

In Sec.~\ref{sec:mgf}, we have shown a method to evaluate the moment generating function of the form [see Eq.~\eqref{cond-avg}]
\begin{align}
    G(\lambda,t) = \langle e^{-\lambda \int_0^t~ds~a^2(s)}\rangle\ , \label{gen-fun-t-0}
\end{align}
where the angular brackets $\langle \cdots \rangle$ indicate the average over trajectories of $a(t)$~\eqref{a-eqn-2}. Comparing with Eq.~\eqref{cond-avg}, we find $q(u)=\lambda$, and the lower limit of the integration to be $s=0$. This implies that the propagator in Eq.~\eqref{Q-full} is  $p(a_s,0|0,0)=\delta(a_s)$. Thus, the moment generating function becomes
\begin{align}
    G(\lambda,t) = Q_{\rm full}(0) = Q(0,0) = 1/\sqrt{y(0)}\ , \label{Glt}
\end{align}
where the right-most equality comes from the solution $Q(a_s,s)$~\eqref{Gau-ans}. 

We now show how to evaluate $y(0)$. Given $q(u)=\lambda$, and the terminal conditions $y(t)=1$ and $y'(t)=0$ [see Sec.~\ref{sec:mgf}], we obtain the solution of the differential equation~\eqref{y-diff}. It is given by
\begin{align}
    y(u) = \cosh[\sqrt{2\lambda}(t-u)]\ .
\end{align}
Substituting $y(0)$ in Eq.~\eqref{Glt} we obtain
\begin{align}
    G(\lambda,t) = \dfrac{1}{\sqrt{\cosh(\sqrt{2\lambda}t)}}\ .\label{Glambdat}
\end{align}

The probability density function $p(Z,t)$ is obtained by inverting the Laplace transform
\begin{align}
    p(Z,t) = \mathcal{L}^{-1} [G(\lambda,t)]\ ,
\end{align}
yielding~(see Appendix E in~\cite{Urna}) 
\begin{align}
    p(Z,t)
    =\dfrac{t}{2\sqrt{\pi Z^3}}\sum_{n=0}^{\infty}(-1)^n \dfrac{4 n + 1}{2^{2n}} \begin{pmatrix}
    2 n \\
    n
    \end{pmatrix} e^{-\frac{t^2 (4n+1)^2}{8Z}}\ .
\end{align}

Now, using $X(t) = X_0 - \sigma_0 t -\sigma Z(t)$ for $Z>0$~\eqref{y-eqn}, and $X = \ln x$,  we obtain the distribution of $x(t)$ as 
\begin{align}
    P(x,t|x_0)& =\dfrac{ \Theta[-\ln x +X_0-t \sigma_0)]\sqrt{\sigma}t}{2x\sqrt{\pi (X_0 -\sigma_0 t - \ln x)^3}}\sum_{n=0}^{\infty}(-1)^n \dfrac{4 n + 1}{2^{2n}} \begin{pmatrix}
    2 n \\
    n
    \end{pmatrix} e^{-\frac{\sigma t^2 (4n+1)^2}{8(X_0 -\sigma_0 t - \ln x)}}\ ,\label{ana-px-d-0}
\end{align}
where $\Theta(\cdot)$ is the Heaviside theta function and $X_0 = \ln x_0$. Notice that the above result holds for $x>0$. However, the solution can be extended to $x<0$ simply by replacing $x\to |x|$.

\begin{figure}
    \centering
    \includegraphics[width=\linewidth]{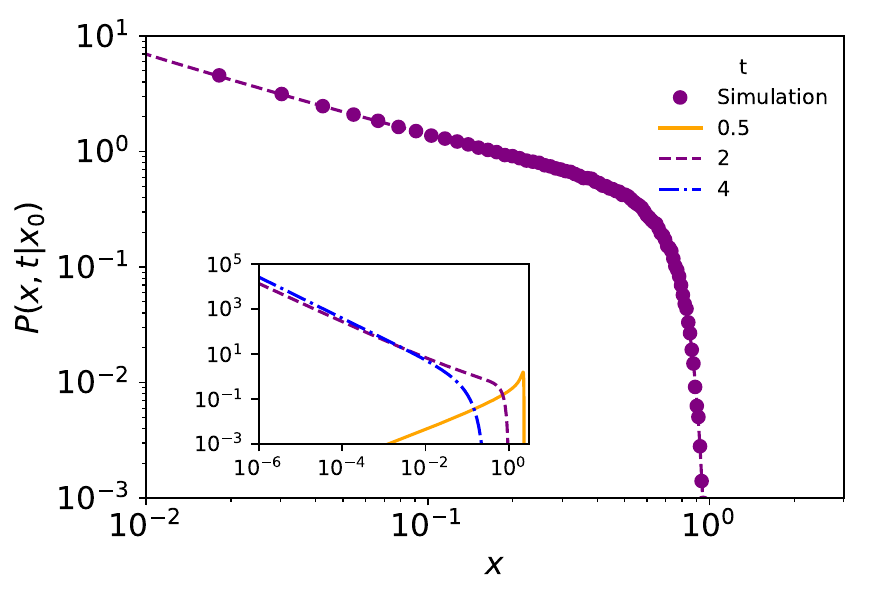}
    \caption{Comparison of the analytical probability density function~\eqref{ana-px-d-0} with the data obtained from the numerical simulations~\eqref{xk-eqn-3} in the absence of thermal fluctuations at time $t=2$. Lines: Analytical result [by summing $10^3$ terms in the summation~\eqref{ana-px-d-0}]. Circles: Numerical simulations performed using the time discretization $dt=10^{-3}$, and number of realizations $10^6$.  The other fixed parameters are strength of the static stiffness $\sigma_0=0.5$, random stiffness $\sigma=3$, the initial position $x_0 = 3$, and the time $t=2$. Inset: Analytical results for $t=0.5, 2, 4$.}
    \label{t-eq0-plot}
\end{figure}

Figure~\ref{t-eq0-plot} shows excellent agreement between the analytical result~\eqref{ana-px-d-0} and the data obtained from the numerical simulation of the Langevin equations~\eqref{xk-eqn-3}. In the inset, we plot the analytical probability density function for three different times. Its behavior as a function of time can be intuitively understood as follows. Since $a(t)$ in Eq.~\eqref{a-eqn18} is a Brownian motion, the stiffness $k(t) = a^2(t)$ increases with time on average. Moreover, the probability of observing higher stiffness increases with time (see Fig.~\ref{fig:pkt}).  Therefore, as the observation time increases, the probability of finding the particle close to the origin increases.

\section{Temperature $T\neq 0$}
\label{sec:gen_tp}
In the previous section, we investigated the positional fluctuations due to the fluctuating stiffness of the trap, in the absence of a heat bath. Now, we investigate the influence of additional thermal fluctuations. Therefore, we consider the full model~\eqref{main-xk-eqn-2} to compute the positional moments. We further rescale the position $x/\sqrt{Dt_k}\to x$ so that the equations of motion read as
\begin{subequations}
\label{main-xk-eqn-4}
\begin{align}
    \dot x(t) &= - [\sigma_0 + \sigma k(t)] x(t) + \sqrt{2}\eta(t) \ , \label{x-eqn-4}\\
     \dot a(t) &= \xi(t)\label{a-eqn-4} \ .
\end{align}
\end{subequations}

To compute the positional fluctuations, we follow the techniques discussed in Ref.~\cite{Gupta_2025-SHT}, where we considered a trapped particle with exponentially correlated Gaussian distributed fluctuating stiffness. We write the Fokker-Planck equation for each realization of $k(t)$
\begin{align}
    \dfrac{\partial p(x,t;k(t))}{\partial t}  = \dfrac{\partial }{\partial x} \big[(\sigma_0 + \sigma k(t)]~x~p(x,t;k(t))\big] +\dfrac{\partial^2 p(x,t;k(t))}{\partial x^2} \ , \label{FPE}
\end{align}
for the initial condition $p(x,0;k_0) = \delta(x-x_0)$ for all $k_0\equiv a^2(0)=0$. The solution of this Fokker-Planck equation, $p(x,t;k(t))$, is then averaged over an ensemble of trajectories of $k(t)$ emanating from $k_0$. This gives the distribution of the position of the particle as follows:
\begin{align}
    P(x,t|x_0) = \langle p(x,t;k(t))\rangle_{\{k(t)\}} \ .\label{ann-avg}
\end{align}

To solve the Fokker-Planck equation~\eqref{FPE} for a given realization of $k(t)$, we use the method of subordination~\cite{MOS-1,MOS-2}, i.e., we use the substitution 
\begin{align}
    p(x,t;k(t)) =  f(t)~\mathcal{P}(z(x,t), \tau(t))\ , \label{p-factor}
\end{align}
for
\begin{subequations}
\label{fzt}
\begin{align}
    f(t) &\equiv  e^{\sigma_0t+\sigma\int_0^t~ds~k(s)}\ ,\label{fteqn}\\
    z(x,t) &\equiv x  f(t)\ \label{eqn-zt},\\
    \tau(t) &\equiv \int_0^t~ds~ f^2(s) \label{tau} \ ,
\end{align}
\end{subequations}
in Eq.~\eqref{FPE}.
This translates the Fokker-Planck equation~\eqref{FPE} to a diffusion equation with stochastic time $\tau$~\eqref{tau}
\begin{align}
    \partial_\tau \mathcal{P}(z,\tau) =\partial_z^2 \mathcal{P}(z,\tau)\ , \label{w-diff}
\end{align}
with the initial condition $\mathcal{P}(z,0) = \delta(z - x_0)$, since $\tau(0)=0$ and $z(x,0)=x_0$. 
The solution of this diffusion equation~\eqref{w-diff} can be obtained using the Fourier-transform
\begin{align}
    \mathcal{P}(z,\tau)
    =\dfrac{1}{2\pi}\int_{-\infty}^{+\infty}~dp~e^{- p^2 \tau} e^{ip(z-x_0)}=\dfrac{1}{2\pi}\int_{-\infty}^{+\infty}~dp~e^{-ipx_0}e^{ipxf(t)}~ e^{- p^2\tau(t)} \ , \label{wsol}
\end{align}
where $p$ is the conjugate variable with respect to $z$, and we have substituted $z$ from Eq.~\eqref{eqn-zt}.  

Substituting $\mathcal{P}(z,\tau)$~\eqref{wsol} into Eq.~\eqref{p-factor} and 
writing the annealed average~\eqref{ann-avg} over the ensemble of trajectories of $k(t)$, we have
\begin{align}
       P(x,t|x_0)=\bigg\langle \int_{-\infty}^{+\infty}~~dp~f(t)\dfrac{e^{-ipx_0}}{2\pi}~ e^{- p^2 \tau(t)} e^{ipxf(t)}\bigg\rangle_{\{k(t)\}} \ . \label{P-full-wo-avg}
\end{align}

It turns out that calculating the above average to obtain $P(x,t|x_0)$ is difficult. This is because of the non-linear terms that are present inside the average, namely $f(t)$ and $\tau(t)$. Nevertheless, the above expression provides a way to obtain positional moments of any order $n$. To this end, we multiply $x$ on both sides of Eq.~\eqref{P-full-wo-avg} and integrate over $x$. This gives 
\begin{align}
    \langle x^n \rangle & = \sum _{q=0, q\in \mathbb{Z}^+}^{n/2} \binom{n}{2q}\frac{4^q}{\sqrt{\pi}}~\Gamma\left(\frac{1+2q}{2}\right)  x_0^{n-2q}\langle f^{-n} \tau^{q}  \rangle \ ,\label{nth-mom}
\end{align}
where the right-most average turns out to be~\cite{Gupta_2025-SHT}
\begin{align}
    \langle f^{-n}(t) \tau^{q}(t) \rangle 
      =  e^{-n \sigma_0 t} \bigg(\prod_{j = 1}^q~\int_0^t~ds_j\bigg)~e^{2 \sigma_0 \sum_{\ell = 1}^q s_\ell}\bigg\langle e^{-\sigma \big[n \int_0^t~ds~k(s) - 2 \sum_{m=1}^q\int_0^{s_m}~da_m~k(a_m)\big]}  \bigg\rangle \ .\label{ftau-first}
\end{align}
In the following subsections, we explicitly calculate the first four moments using the expressions~\eqref{nth-mom} and \eqref{ftau-first}.
\subsection{First moment}
For the first positional moment ($n=1$), we find from Eq.~\eqref{nth-mom}
\begin{subequations}
    \begin{align}
    \langle x \rangle &= x_0 \langle f^{-1}(t)\rangle=x_0  e^{-\sigma_0 t} \langle e^{-\sigma \int_0^t~ds~k(s)} \rangle = x_0  e^{-\sigma_0 t} G(\sigma, t)\label{first-mom-1} \\ 
    &=\dfrac{x_0  e^{-\sigma_0 t} }{\sqrt{\cosh(\sqrt{2\sigma}t)}}\ , \label{first-mom}
\end{align}
\end{subequations}
where in Eqs.~\eqref{first-mom-1} and \eqref{first-mom}, we used Eqs.~\eqref{gen-fun-t-0} and \eqref{Glambdat}, respectively. 

The leading contribution in the long-time limit
\begin{align}
   \langle x \rangle \approx \sqrt{2}x_0 e^{-(\sigma_0 +\sqrt{\sigma/2})t} \ , \label{first-mom-scaling}
\end{align}
which goes to $0$ as $t \to \infty$. Therefore, a particle in a stiffness-fluctuating trap, starting from $x_0 \neq 0$, will have a zero mean position in the long-time limit.

\subsection{Second moment}
For the second moment $\langle x^2 \rangle$, Eq.~\eqref{nth-mom} gives us 
\begin{align}
    \langle x^2\rangle &= x_0^2 \langle f^{-2}(t)\rangle + 2 \langle f^{-2}(t)\tau(t)\rangle \\
    &=x_0^2  e^{-2\sigma_0 t} \langle e^{-2\sigma \int_0^t~ds~k(s)} \rangle + 2 \int_0^t~e^{-2\sigma_0(t-s)}\langle e^{-2\sigma \int_s^t~du~k(u)}\rangle \ ,
\end{align}
where the first term can be solved using Eqs.~\eqref{gen-fun-t-0} and \eqref{Glambdat}. Then, we have 
\begin{align}
    \langle x^2\rangle 
    &=\dfrac{x_0^2  e^{-2\sigma_0 t}}{\sqrt{\cosh(\sqrt{4\sigma}t)}} + 2 \int_0^t~e^{-2\sigma_0(t-s)}\langle e^{-2\sigma \int_s^t~du~k(u)}\rangle \ .\label{sec-mom}
\end{align}
We now discuss how to compute the average in the second term on the right-hand side of Eq.~\eqref{sec-mom}. Following Sec.~\ref{sec:mgf}, this average turns out to be $Q_{\rm full}(s)$~\eqref{fin-avg} with $q(u) = 2\sigma$ for $s\leq u\leq t$. To obtain $Q_{\rm full}(s)$, we need to solve the differential equation~\eqref{y-diff} for $y(s)$ given the terminal conditions below Eq.~\eqref{y-diff}. This gives 
\begin{align}
    y(s) = \cosh[\sqrt{4\sigma}(t-s)]\ .
\end{align}
Substituting $y(s)$ in Eq.~\eqref{sub-alp} gives $\alpha(s)$, which we substitute in Eq.~\eqref{beta-dff} into $\beta(s)$, yielding
\begin{subequations}
    \begin{align}
    \alpha(s) = \sqrt{\sigma}\tanh[\sqrt{4\sigma}(t-s)]\ ,\\
    \beta(s) = -\dfrac{1}{2}\ln \cosh[\sqrt{4\sigma}(t-s)] \ .
\end{align}
\end{subequations}
These quantities are then substituted in Eq.~\eqref{Gau-ans} to compute $Q(a_s,s)$. Using Eq.~\eqref{fin-avg}, we then obtain
\begin{align}
 Q_{\rm full}(s) 
 =\dfrac{1}{\sqrt{\cosh(\sqrt{4\sigma}(t-s)) + 2 s \sqrt{\sigma}\sinh(\sqrt{4\sigma}(t-s))}}\ .
 \end{align}
Therefore, the second moment in the integral form~\eqref{sec-mom} is given by
\begin{align}
    \langle x^2\rangle 
    &=\dfrac{x_0^2  e^{-2\sigma_0 t}}{\sqrt{\cosh(\sqrt{4\sigma}t)}} + 2 \int_0^t~ds~\dfrac{e^{-2\sigma_0 s}}{\sqrt{\cosh(\sqrt{4\sigma}s) + 2 (t-s) \sqrt{\sigma}\sinh(\sqrt{4\sigma}s)}} \ .\label{sec-mom-2}
\end{align}
In the long-time limit, the first term on the right-hand side approaches $\sqrt{2}x_0^2e^{-(2\sigma_0 +\sqrt{\sigma})t}$, which goes to zero as $t\to \infty$. Since the integrand is dominated by small-$s$ contributions for a given $t$, the quantity inside the square-root in the integrand is dominated by the term proportional to $t-s\approx t$ in the long-time limit. Thus, the second positional moment in the long-time is given as 
\begin{align}
    \langle x^2\rangle 
    &= C_2(\sigma_0,\sigma)~t^{-1/2}  \ ,\label{sec-mom-2-approx}
\end{align}
where we defined the coefficient
\begin{align}
    C_2(\sigma_0,\sigma) \equiv \dfrac{\sqrt{2}}{\sigma^{1/4}} \int_0^\infty~ds~\dfrac{e^{-2\sigma_0 s}}{\sqrt{\sinh(\sqrt{4\sigma}s)}} = \frac{\sqrt{\pi } \Gamma \left(\frac{\sigma_0}{2 \sqrt{\sigma }}+\frac{1}{4}\right)}{2 \sigma ^{3/4} \Gamma \left(\frac{\sigma_0}{2 \sqrt{\sigma }}+\frac{3}{4}\right)}\ .
\end{align}

Therefore, the positional variance [using Eqs.~\eqref{first-mom-scaling} and~\eqref{sec-mom-2-approx}] decreases as $t^{-1/2}$ in the long-time limit. 

\subsection{Third moment}
\label{sec:third-mom}
For the third moment ($n=3$), Eq.~\eqref{nth-mom} gives
\begin{align}
    \langle x^3 \rangle &= x_0^3 \langle f^{-3}(t)\rangle  + 6x_0 \langle f^{-3}(t)\tau(t)\rangle \\
    &=x_0^3 e^{-3 \sigma_0 t}\langle e^{-3\sigma\int_0^t~ds~k(s)}\rangle + 6x_0 e^{- \sigma_0 t} \int_0^t~ds~e^{-2\sigma_0(t-s)}\langle e^{-\sigma\int_0^t~ds~k(s)}e^{-2\sigma\int_s^t~du~k(u)}\rangle \ , \label{eq:48}
\end{align}
where the first term can be evaluated using Eqs.~\eqref{gen-fun-t-0} and \eqref{Glambdat}. This yields
\begin{align}
    \langle x^3 \rangle 
    =\dfrac{x_0^3  e^{-3 \sigma_0 t}}{\sqrt{\cosh(\sqrt{6\sigma}t)}} + 6x_0 e^{- \sigma_0 t} \int_0^t~ds'~e^{-2\sigma_0(t-s')}\langle e^{-\int_0^t~du~q_1(u)~a^2(u)}\rangle \ ,\label{x3-avg}
\end{align}
where 
\begin{align}
    q_1(u) =\begin{cases}
    \sigma &\qquad 0\leq u< s'\\
    3\sigma&\qquad s'\leq u\leq t
    \end{cases}\ .\label{qu-1}
\end{align}

Then, the average on the right-hand side of Eq.~\eqref{eq:48}, by definition, is given by~\eqref{fin-avg}
\begin{align}
    Q_{\rm full}(0) = Q(0,0)\ .\label{Qs-Qf0} 
\end{align}
To proceed, we start from $Q(a_s,s)$~\eqref{cond-avg}. 
This will lead us to the differential equation~\eqref{y-diff}. Given the terminal conditions [below Eq.~\eqref{y-diff}], we first solve the differential Eq.~\eqref{y-diff} in the interval $u\in[s',t]$ with $q_1(u)=3\sigma$ [Eq.~\eqref{qu-1}]. Then, we solve the interval $u\in [s,s']$ with $q_1(u)=\sigma$ [Eq.~\eqref{qu-1}]. The complete solution is constructed using the matching conditions $y(u)$ and its derivative $y'(u)$ at $u=s'$. Again, this means that we are solving the differential Eq.~\eqref{y-diff} {\it backwards} in time.

We assume that the general solution of Eq.~\eqref{y-diff} in the interval $u\in [s',t]$ is 
\begin{align}
    y(u) = A\cos[\omega u] + B \sin[\omega u]\ . \label{gen-sol}
\end{align}
Using the terminal conditions $y(t)=1, \dot y(t)=0$ [see below Eq.~\eqref{y-diff}], the solution~\eqref{gen-sol} becomes
\begin{align}
    y(u) = \cos[\omega (t-u)]\ . \label{gen-sol-2}
\end{align}

Substituting the solution~\eqref{gen-sol-2} in the differential equation~\eqref{y-diff}
\begin{align}
    \omega^2 + 6\sigma = 0 \implies \omega = \pm i\sqrt{6\sigma}\ .
\end{align}
Then, the solution~\eqref{gen-sol-2} becomes
\begin{align}
    y(u) = \cosh[\sqrt{6\sigma}(t-u)]\ . \label{right-sol}
\end{align}

Similarly, the general solution for the interval $u\in [s,s']$ satisfies the differential equation~\eqref{y-diff}
\begin{align}
    y(u) = C\cosh[\sqrt{2\sigma} u] + D \sinh[\sqrt{2\sigma} u]\ . \label{left-sol}
\end{align}

The coefficients $C$ and $D$ [using Eqs.~\eqref{right-sol} and \eqref{left-sol}] can be obtained by using the matching conditions at $u=s'$. Substituting these $C$ and $D$, we find the full solution for $s\leq u\leq t$. This is given by
\begin{align}
   y(u)=\cosh \left(\sqrt{6\sigma} (t-s')\right) \cosh \left(\sqrt{2\sigma} (s'-u)\right) +  \sqrt{3}\sinh \left(\sqrt{6\sigma} (t-s')\right) \sinh \left(\sqrt{2\sigma} (s'-u)\right)\ .\label{y-s-to-t}
\end{align}

Using the Gaussian ansatz~\eqref{Gau-ans} and 
\begin{align}
    \beta(s) = \dfrac{1}{2}\int_s^t~du~\dfrac{d}{du}\ln y(u) = -\dfrac{1}{2}\ln y(s)\ ,
\end{align}
where we used the fact that $y(t)=1$, we obtained the conditional average~\eqref{cond-avg} 
\begin{align}
    Q(a,s)=\dfrac{e^{-\alpha(s)a^2}}{\sqrt{y(s)}}\ .
\end{align}
However, as we discussed above in Eq.~\eqref{Qs-Qf0}, the average on the right-hand side of Eq.~\eqref{x3-avg} is given by 
\begin{align}
    Q_{\rm full}(0)= \dfrac{1}{\sqrt{y(0)}}\ . 
\end{align}

This implies that the third moment~\eqref{x3-avg} in the integral form is given by
\begin{align}
    \langle x^3 \rangle 
    =\dfrac{x_0^3  e^{-3 \sigma_0 t}}{\sqrt{\cosh(\sqrt{6\sigma}t)}} + 6x_0 e^{- \sigma_0 t} \int_0^t~ds'~\dfrac{e^{-2\sigma_0(t-s')}}{\sqrt{y(0)}}\ ,\label{x3-avg-2}
\end{align}
where $y(0)$ is obtained from Eq.~\eqref{y-s-to-t}.

In the long-time limit, leading order contribution to the third moment is
\begin{align}
  \langle x^3 \rangle \approx \sqrt{2}x_0^3 e^{-(3\sigma_0 + \sqrt{6\sigma}/2)t} +  6\sqrt{2}x_0 e^{-(\sigma_0 +\sqrt{\sigma/2})t} \int_0^t~ds~\dfrac{e^{-2\sigma_0 s}}{\sqrt{\cosh(\sqrt{6\sigma}s) + \sqrt{3}\sinh(\sqrt{6\sigma}s)}}\ . \label{3rd-mom}
\end{align}
In this limit ($t\to\infty$), the integral on the right-hand side~\eqref{3rd-mom} converges by extending the upper integration limit to $t=\infty$. Therefore, the third moment approaches zero exponentially.

\subsection{Fourth moment}
Finally, we present the computation of the fourth positional moment by substituting $n=4$ in Eq.~\eqref{nth-mom}, yielding
\begin{align}
\langle x^4 \rangle 
&= x_0^{4}\langle f^{-4}(t)\rangle + 12x_0^2~\langle f^{-4}(t) \tau(t)  \rangle + 12\langle f^{-4}(t) \tau^{2}(t)  \rangle\ ,
\end{align}
where the first term on the right-hand side is again obtained using Eqs.~\eqref{gen-fun-t-0} and \eqref{Glambdat}. Then, the fourth moment reads
\begin{align}
\langle x^4 \rangle 
&= \dfrac{x_0^{4}e^{-4 \sigma_0 t}}{\sqrt{\cosh(\sqrt{8\sigma} t)}} + 12x_0^2~e^{-2\sigma_0 t}\int_0^t~ds'~e^{-2\sigma_0 (t-s')} \underbrace{\langle e^{-\int_0^t~du~q_2(u)~k(u)}\rangle}_{\mathcal{A}_1(t,s')} + \nonumber\\
&+ 12 \int_0^t~ds\int_0^t~ds'~e^{-2\sigma_0(t-s)} e^{-2\sigma_0(t-s')} \underbrace{\langle e^{-2\sigma\int_{s}^{t}~du~q_3(u)~k(u)}\rangle}_{\mathcal{A}_2(t,s,s')}\ , \label{x4-mom}
\end{align}
where we identified 
\begin{subequations}
    \begin{align}
    q_2(u) = \begin{cases}
        2\sigma \qquad 0\leq u< s'\\
        4\sigma \qquad s'\leq u\leq t
    \end{cases}\ ,\\
    q_3(u) = \begin{cases}
        2\sigma \qquad s\leq u< s'\\
        4\sigma \qquad s'\leq u\leq t
    \end{cases}
     .
\end{align}
\end{subequations}

Similarly to what is shown in the previous subsection~\ref{sec:third-mom}, we can show that 
\begin{subequations}
\label{A12}
\begin{align}
   \mathcal{A}_1(t,s') &= \dfrac{1}{\sqrt{y(0)}}\ ,\\
    \mathcal{A}_2(t,s,s') &= Q_{\rm full}(s) = \dfrac{1}{\sqrt{y(s)}\sqrt{1 +2s \alpha(s)}}\ , 
\end{align}    
\end{subequations}
where $y(u)$ is given by Eq.~\eqref{y-s-to-t} [after replacing $q_1(u)$ with $q_3(u)$]:
\begin{align}
   y(u)=\cosh \left(\sqrt{8\sigma} (t-s')\right) \cosh \left(\sqrt{4\sigma} (s'-u)\right) +  \sqrt{2}\sinh \left(\sqrt{8\sigma} (t-s')\right) \sinh \left(\sqrt{4\sigma} (s'-u)\right)\ ,
\end{align}
and $\alpha(s)$ can be obtained from Eq.~\eqref{sub-alp}. Thus, together with $\mathcal{A}_{1,2}$~\eqref{A12}, we can obtain the fourth positional moment~\eqref{x4-mom}.
\begin{figure}
    \centering
    \includegraphics[width=0.8\linewidth]{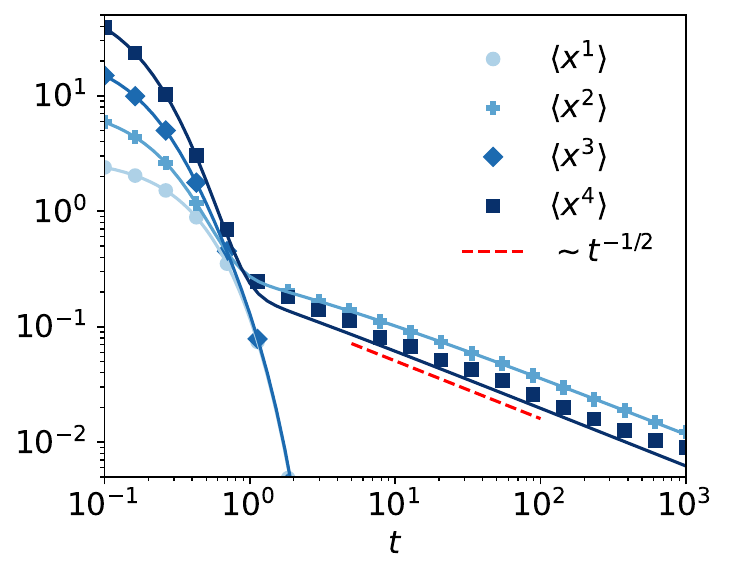}
    \caption{Comparison of analytical (lines) and numerical simulations data (symbols) of first four positional moments as a functions of time $t$. Analytical results: Eqs.~\eqref{first-mom}, \eqref{sec-mom}, \eqref{x3-avg-2}, \eqref{x4-mom}. Numerical simulations were performed for time discretization $dt=10^{-5}$ and number of realizations $10^5$. The other fixed parameters are strength of the static $\sigma_0=2$, the random stiffness $\sigma=5$, and the initial position $x_0 = 3$. }
    \label{fig:moments}
\end{figure}
In the long-time limit, the leading contribution to the fourth moment is 
\begin{align}
    \langle x^4 \rangle \sim C_1 e^{-(4\sigma_0 + 2 \sqrt{\sigma})t} + C_2 e^{-(2\sigma_0 + \sqrt{\sigma})t} + C_3 t^{-1/2} \, \label{4th-mom-scaling}
\end{align}
where $C_{1,2,3}$ are constants coming from three terms in Eq.~\eqref{x4-mom}. Therefore, in the long-time limit, the fourth moment scales as $t^{-1/2}$ as the second moment. 
Finally, we remark that, as shown above, higher moments can be computed in the same fashion.

Figure~\ref{fig:moments} demonstrates a good agreement between the analytical expressions for the first four positional moments [Eqs.~\eqref{first-mom}, \eqref{sec-mom}, \eqref{x3-avg-2}, and \eqref{x4-mom}] and the numerical simulation results obtained from the Langevin equations~\eqref{main-xk-eqn-4}. As evident from the figure, at long times, the first and third moments decay exponentially to zero, whereas the second and fourth moments decay as $t^{-1/2}$. This long-time decaying behavior of all the moments arises because the stiffness \(k(t)\) is modeled as the square of a Brownian motion, implying that its mean value increases with time~\eqref{eq:k-avg}. Furthermore, Fig.~\ref{fig:pkt} shows that the probability distribution of the stiffness shifts progressively toward larger values as time evolves. The resulting increase in confinement suppresses positional fluctuations, causing all moments to decay over time.

\section{Thermodynamics}
\label{sec:thermo}
In this section, we discuss the thermodynamic properties of the system. In particular, we examine how much work is applied due to time-dependent fluctuations in the stiffness of the trap and how much heat the particle exchanges with the environment.

{\color{black}Starting from the internal energy} $U(x;k(t))$ of the system, the rates of external work and heat flow along a stochastic trajectory can be identified by computing the total derivative of the internal energy~\cite{sekimoto2010stochastic}. This gives
\begin{align}
    \dfrac{d U(x;k(t))}{d t} = \underbrace{\dfrac{\partial U(x;k(t))}{\partial k}\dot k}_{\dot w} + \underbrace{\dfrac{\partial U(x;k(t))}{\partial x} \dot x}_{\dot q} \ ,\label{FLT}
\end{align}
where the first and second terms, respectively, on the right-hand side are the rates of work $\dot w$ and heat $\dot q$ along a stochastic trajectory. 

Therefore, the total work performed on the particle and the heat exchanged by the particle with the bath, up to time $t$, are given, respectively, by
\begin{align}
    w &= \int_0^t~ds~\dfrac{\partial U(x;k(s))}{\partial k}\dot k\ ,  \label{work-eqn} \\
    q &= \int_0^t~ds~\dfrac{\partial U(x;k(s))}{\partial x} \dot x\ . \label{heat-eqn}
\end{align}
Notice that $w>0$ and $q>0$ correspond, respectively,  to the situation when the work is performed on the system and the heat flows from the bath to the system. 

Substituting the expression for $U(x;k(\tau))$~\eqref{x-eqn-2}, we rewrite the expression for the work~\eqref{work-eqn}
\begin{align}
    w = \dfrac{\sigma}{2}\int_0^t~ds~\dot k x^2\ .\label{work-eqn-2}
\end{align}
This, $w$, is a stochastic quantity due to thermal noise $\eta(t)$~\eqref{x-eqn-2} and fluctuations in the stiffness $k(t)$~\eqref{a-eqn-2}. Thus, to compute the average work, we have to take the average over the trajectories of both thermal noise $\eta(t)$ and stiffness $k(t)$.

To this end, we first average over thermal noise $\eta(t)$ for a given trajectory of $k(t)$, and this gives 
\begin{align}
    \mathcal{W} = \dfrac{\sigma}{2}\int_0^t~ds~\dot{k}(s)~\mathcal{V}(s)\ ,\label{w-eqn3}
\end{align}
where we defined the second positional moment, $\mathcal{V}(t) \equiv \int_{-\infty}^{+\infty}~dx~x^2~p(x,t;k(t))$, for a given trajectory of $k(t)$, and $p(x,t;k(t))$ is the solution of the Fokker-Planck equation~\eqref{FPE}.  We rewrite Eq.~\eqref{w-eqn3} as
\begin{align}
    \mathcal{W} =  \dfrac{\sigma}{2}\int_0^t~ds~\bigg[\dfrac{d [k(s) \mathcal{V}(s)]}{ds}-k(s) \dot {\mathcal{V}}(s)\bigg]\ . \label{w-eqn0}
\end{align}
Taking a time-derivative of both sides of $\mathcal{V}(t) \equiv \int_{-\infty}^{+\infty}~dx~x^2~p(x,t;k(t))$ and using the Fokker-Planck equation~\eqref{FPE}, we obtain the equation of motion for the second positional moment for a given trajectory of $k(t)$
\begin{align}
    \dot{\mathcal{V}}(t) = -2[\sigma_0 + \sigma k(t)]\mathcal{V}(t) + 2\ .\label{eom-v}
\end{align}
Substituting the above equation~\eqref{eom-v} on the right-hand side of Eq.~\eqref{w-eqn0} and then averaging over the trajectories of $k(t)$, we obtain
\begin{align}
    W = \dfrac{\sigma}{2}\int_0^t~ds~\bigg[\dfrac{d \langle k(s) \mathcal{V}(s)\rangle}{ds}-2\langle k(s)\rangle + 2\sigma_0 \langle k(s)\mathcal{V}(s)\rangle +2 \sigma \langle k^2(s)\mathcal{V}(s)\rangle \bigg]\ . \label{w-eqn4}
\end{align}
To simplify the integrand on the right-hand side of the above Eq.~\eqref{w-eqn4}, we calculate the equation of motion of the mixed moments $M_n(t)\equiv \langle a^n(t) \mathcal{V}(t)\rangle$, where $k(t)= a^2(t)$. This is given by (see~\ref{sec:mixed} for details)
\begin{align}
\frac{d}{dt}\langle a^n(t) \mathcal{V}(t)\rangle=2\langle a^n\rangle- 2\sigma_0 \langle a^n \mathcal{V}\rangle- 2\sigma \langle a^{n+2}\mathcal{V}\rangle + \frac{1}{2}n(n-1)\langle a^{n-2}\mathcal{V}\rangle\ .
\end{align}
Substituting $n=2$ in the above equation and rearranging the terms, we recognize the integrand on the right-hand side of Eq.~\eqref{w-eqn4} as
\begin{align}
    \frac{d}{dt}\langle k \mathcal{V}\rangle - 2\langle k\rangle + 2\sigma_0 \langle k \mathcal{V}\rangle +  2\sigma \langle k^2\mathcal{V}\rangle = \langle \mathcal{V}\rangle\ .\label{re-arr}
\end{align}
Therefore, the average work becomes
\begin{align}
    W = \dfrac{\sigma}{2}\int_0^t~ds~ \langle \mathcal{V}(s)\rangle \ ,\label{w-fin}
\end{align}
where $\langle \mathcal{V}(s)\rangle \equiv \langle x^2\rangle$ is given in Eq.~\eqref{sec-mom-2}. It is not straightforward to analytically evaluate the integral~\eqref{w-fin}; nevertheless, one can calculate the integral numerically. Furthermore, it is interesting to analyze average work~\eqref{w-fin} in the long-time limit. Since $\langle \mathcal{V}(s)\rangle$ scales as $s^{-1/2}$~\eqref{sec-mom-2-approx} in the long time limit, the average work grows {\color{black}nonlinearly as}
\begin{align}
    W \sim t^{1/2}\ . \label{eq:work-scaling}
\end{align}

Next, we compute the average heat exchanged by the particle with the heat bath. This is obtained from the first law of thermodynamics~\eqref{FLT}
\begin{subequations}
    \begin{align}
    Q &= \langle \Delta U\rangle-W\\
    &= \int_0^t~ds \bigg[\dfrac{\sigma_0}{2} \dfrac{d}{ds}\langle \mathcal{V}(s)\rangle + \dfrac{\sigma}{2} \dfrac{d}{ds}\langle k(s)\mathcal{V}(s)\rangle\bigg] - W\ , \label{eq:Q-exp}
\end{align}
\end{subequations}
where the terms inside the square brackets can be simplified using the equation of motion~\eqref{eom-v}, and we finally arrive at 
\begin{align}
    Q = -\dfrac{1}{4} \bigg[\dfrac{d \langle \mathcal{V}(s)\rangle}{ds}\bigg|_{s=t} - \dfrac{d \langle \mathcal{V}(s)\rangle}{ds}\bigg|_{s=0}\bigg] - \dfrac{\sigma}{2}\int_0^t~ds~ \langle \mathcal{V}(s)\rangle\ .\label{eq:avgq}
\end{align}
Substituting $\langle \mathcal{V}(s)\rangle \equiv \langle x^2\rangle$~\eqref{sec-mom-2} in the first term on the right-hand side (i.e., the internal energy change) yields
\begin{align}
\Delta U = -\dfrac{1}{4} \bigg[\dfrac{d \langle \mathcal{V}(s)\rangle}{ds}\bigg|_{s=t} - \dfrac{d \langle \mathcal{V}(s)\rangle}{ds}\bigg|_{s=0}\bigg] = -\dfrac{1}{4}[\mathcal{U}_1(t) - \mathcal{U}_1(0) + \mathcal{U}_2(t)]\ ,
\end{align}
where we defined 
\begin{subequations}
\begin{align}
    \mathcal{U}_1(t)&\equiv -\frac{x_0^2 e^{-2 \sigma_0 t} \left[2 \sigma_0 \cosh \left(2 \sqrt{\sigma } t\right)+\sqrt{\sigma } \sinh \left(2 \sqrt{\sigma } t\right)\right]}{\cosh ^{\frac{3}{2}}\left(2 \sqrt{\sigma } t\right)}\ ,\\
    \mathcal{U}_2(t) &= -\int_0^t~ds~\frac{2 e^{-2 \sigma_0 (t-s)} \left[\sqrt{\sigma } (4 s \sigma_0+1) \sinh \left(2 \sqrt{\sigma } (t-s)\right)+2 (s \sigma +\sigma_0) \cosh \left(2 \sqrt{\sigma } (t-s)\right)\right]}{\left[2 s \sqrt{\sigma } \sinh \left(2 \sqrt{\sigma } (t-s)\right)+\cosh \left(2 \sqrt{\sigma} (t-s)\right)\right]^{3/2}}\ .
\end{align}    
\end{subequations}
In the long-time limit, these terms behave as
\begin{subequations}
 \begin{align}
    \mathcal{U}_1(t)&\approx C_1~e^{-t(2\sigma_0 + \sqrt{\sigma})}\ ,\\
        \mathcal{U}_2(t)&\approx C_2~t^{-1/2}\ ,
\end{align}   
\end{subequations}
where $C_{1,2}$ are constants. 
Thus, in the long-time limit, the change in the internal energy reaches a stationary value 
\begin{align}
    \Delta U = \dfrac{\mathcal{U}_1(0)}{4} = -\dfrac{\sigma_0 x_0^2}{2}\ ,\label{eq:ssdu}
\end{align}
which arises from integrating the first term of the square bracket~\eqref{eq:Q-exp} by noticing that $\langle x^2\rangle_{t\to \infty}$ decays to zero as $t^{-1/2}$ [Eq.~\eqref{sec-mom-2-approx}]. 
Therefore, in this limit, the average heat~\eqref{eq:avgq}, similar to the average work~\eqref{eq:work-scaling}, also grows nonlinearly. We stress again that both quantities scale nonlinearly in time, revealing the nonstationary character of this nonequilibrium system.

\begin{figure}
    \centering
    \includegraphics[width=0.8\linewidth]{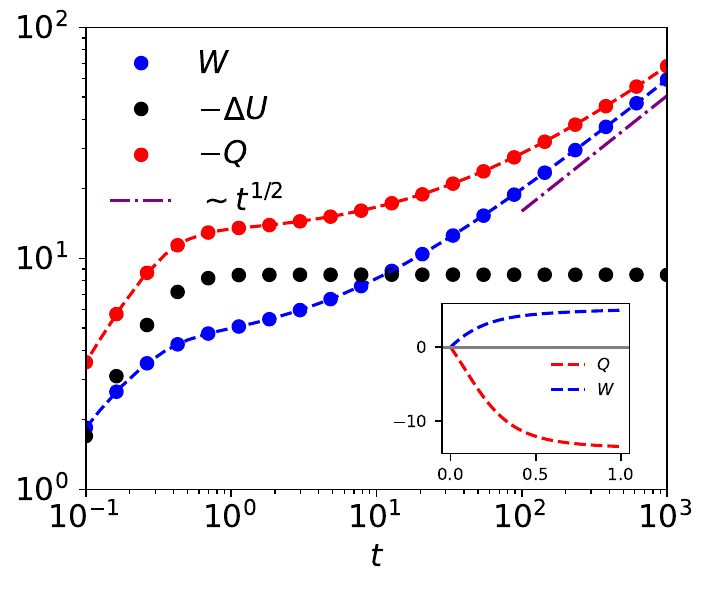}
    \caption{Average work $W$ (blue), average heat dissipated in the bath $-Q$ (red), and the negative change in the average internal energy $-\Delta U$ (black), each as a function of time $t$. Symbols: Numerical simulations. Dashed lines: Analytical results~\eqref{w-fin} and \eqref{eq:avgq}. The parameters are $\sigma_0=2$, $\sigma = 5$, and $x_0 = 3$. For numerical simulations, we take $dt = 10^{-5}$ and number of realizations $10^{5}$. Inset: Analytical results for $W$ and $Q$.  }
    \label{fig:WQU}
\end{figure}

Figure~\ref{fig:WQU} shows excellent agreement between the theoretical prediction of the average work~\eqref{w-fin} and the average heat~\eqref{eq:avgq}, and the numerical Langevin simulations. Furthermore, in the long-time limit, we confirm the net change in the internal energy $\Delta U = -\sigma_0 x_0^2/2$~\eqref{eq:ssdu} as well as the nonlinear scaling laws for work~\eqref{eq:work-scaling} and heat.

The results in Fig.~\ref{fig:WQU} can be understood intuitively as follows. At time $t=0$, the particle is located at $x_0=3$, while the trap center is at $x=0$. At this initial stage, the trap is static; therefore, the average work is zero [Fig.~\ref{fig:WQU}(inset)]. However, since the particle is initially displaced from the trap center, it relaxes toward lower-energy configurations by dissipating heat into the thermal bath; consequently, the average heat flow is negative [Fig.~\ref{fig:WQU}(inset)].

As time increases, fluctuations in the trap stiffness lead to an increase in the work performed on the system. Moreover, the stochastic stiffness is a non-stationary process [see Eq.~\eqref{eq:pkt} for its probability distribution], whose mean value increases linearly with time, $\langle k(t)\rangle = t$ [Eq.~\eqref{eq:k-avg}]. Therefore, in the long-time limit, the average work is not expected to grow linearly over time [Fig.~\ref{fig:WQU}], in contrast to the case where fluctuations of the stiffness are described by a stationary process~\cite{Gupta_2025-SHT}.

 Once the transient effects associated with the initial condition have disappeared, the system reaches a regime where the average internal energy change reaches a stationary value of  $-\sigma_0 x_0^2/2$~\eqref{eq:ssdu}. This stationary value corresponds exactly to the energy initially injected into the system by placing the particle at $x_0$. Consequently, according to the first law of thermodynamics, the average heat dissipated by the system balances the average work performed on it~\eqref{FLT}. In other words, the rate of heat dissipation equals the rate at which work is injected into the system. 

\section{Summary}
\label{sum}
In this work, we investigated the dynamics and thermodynamics of a Brownian particle confined in a harmonic trap whose stiffness fluctuates in time, i.e., annealed disorder. Specifically, we considered a stochastic stiffness modeled as the square of a Brownian motion, ensuring that the stiffness remains non-negative while exhibiting temporal fluctuations.

We first analyzed the dynamics in the absence of thermal noise and obtained an exact analytical expression for the probability density function of the particle's position at time $t$. Our results revealed that the probability of finding the particle near the trap center increases with time. This behavior reflects the effect of the randomly increasing confinement generated by the fluctuating stiffness.

We then incorporated thermal fluctuations by coupling the particle to a heat bath and derived exact analytical expressions for the first four positional moments. We found that the odd moments decay exponentially with time, whereas the even moments exhibit a different behavior and decay algebraically as $t^{-1/2}$ in the long-time limit, highlighting the nontrivial influence of stochastic confinement on the particle's position fluctuations.

The temporal fluctuations of the trap stiffness continuously drive the system away from thermal equilibrium. To understand this nonequilibrium process, we investigated the thermodynamic properties of the system within the framework of stochastic thermodynamics. In particular, we analytically calculated the average work performed on the particle by the fluctuating trap and the average heat exchanged between the particle and the surrounding thermal reservoir for all time. In the long time limit, both quantities exhibit asymptotic scaling behavior as $t^{1/2}$. The results presented here assume a fixed initial condition $x_0$. The extension to an initial condition drawn from a distribution $\rho(x_0)$ is straightforward and is obtained by averaging the results over the distribution $\rho(x_0)$.

Our work opens several avenues for future research. A natural extension would be to consider trap stiffnesses generated by higher even powers ($n>2$) of a Brownian process and to investigate how such nonlinear fluctuations affect the statistical and thermodynamic properties of the confined particle. Another interesting direction is to study the case in which the stiffness is given by the square of an Ornstein-Uhlenbeck process, thereby introducing positive fluctuations with a finite correlation time. More generally, it would be worthwhile to explore systems in which both the trap location and the trap stiffness fluctuate simultaneously while ensuring the positivity of the stiffness. Such models are expected to exhibit rich nonequilibrium behavior and will be a topic of future discussions. 

We would like to add that our model could be useful for understanding the biodiversity of ecological communities with positively fluctuating metabolic strategies~\cite{zanchetta2025emergenceecologicalstructurespecies}. Finally, our system can be realized in an experiment using Brownian particles in a harmonic trap, whose stiffness can be modulated externally.

\ack
D.G. thanks Luca Cocconi (University of Cambridge), Luís B. Pires (Universidade Federal de Viçosa), and Kainã G. Diniz (Universidade Federal do Rio de Janeiro) for many insightful discussions.
The authors thank Sofia Samaniego (Technische Universität Berlin) for carefully reading the manuscript. D.G. acknowledges the support from the Alexander von Humboldt foundation. \\

\appendix

\section{Evolution of mixed moments: $d/dt \langle a^n(t) V(t)\rangle$}
\label{sec:mixed}
The evolution of $V(t)$ and $a(t)$ is given by
\begin{align}
\dot{\mathcal{V}}(t) &= 2 - 2[\sigma_0 + \sigma a^2(t)]\mathcal{V}(t)\ ,\\
\dot a(t) &= \xi(t)\ .
\end{align}

Then, the joint probability density function $(P(a,\mathcal{V},t)$ satisfies the following Fokker-Planck equation
\begin{align}
\partial_t P(a,\mathcal{V},t)=-\partial_\mathcal{V}\Big( [2 - 2(\sigma_0 + \sigma a^2)\mathcal{V}]\,P \Big) + \frac{1}{2}\partial_a^2 P\ .
\end{align}

We define the mixed moments as follows
\begin{align}
M_n(t) = \langle a^n(t) \mathcal{V}(t) \rangle = \int da\, d\mathcal{V}\, a^n \mathcal{V}\, P(a,\mathcal{V},t)\ .
\end{align}

Taking the time derivative,
\begin{align}
\frac{d}{dt}M_n =\int da\, d\mathcal{V}\, a^n \mathcal{V} \partial_t P(a,\mathcal{V},t)\ .
\end{align}

Substituting the Fokker-Planck equation,
\begin{align}
\frac{d}{dt}M_n=\int da\, d\mathcal{V}\, a^n \mathcal{V} \Big[-\partial_\mathcal{V}\big( [2 - 2(\sigma_0 + \sigma a^2)\mathcal{V}]P \big)+ \frac{1}{2}\partial_a^2 P(a,\mathcal{V},t)\Big]\ .
\end{align}

Performing integration by parts, we obtain
\begin{align}
\frac{d}{dt}\langle a^n \mathcal{V}\rangle=2\langle a^n\rangle- 2\sigma_0 \langle a^n \mathcal{V}\rangle- 2\sigma \langle a^{n+2}\mathcal{V}\rangle + \frac{1}{2}n(n-1)\langle a^{n-2}\mathcal{V}\rangle\ .
\end{align}

We can clearly check that for $n=0$, we get
\begin{align}
\frac{d}{dt}\langle \mathcal{V}\rangle=2- 2[\sigma_0 \langle \mathcal{V}\rangle  + \sigma \langle a^2 \mathcal{V}\rangle]\ .
\end{align}

Hence,
\begin{align}
\langle a^2 \mathcal{V}\rangle=\frac{2 - \dot{\langle \mathcal{V}\rangle} - 2\sigma_0 \langle \mathcal{V}\rangle}{2\sigma}\ .
\end{align}

For $n=2$, we obtain
\begin{align}
\frac{d}{dt}\langle a^2 \mathcal{V}\rangle = 2\langle a^2\rangle - 2\sigma_0 \langle a^2 \mathcal{V}\rangle - 2\sigma \langle a^4 \mathcal{V}\rangle + \langle \mathcal{V}\rangle\ .
\end{align}
Given this, we can obtain 
\begin{align}
   \langle a^4 \mathcal{V}\rangle = \dfrac{1}{2\sigma}\bigg[2\langle a^2\rangle - 2\sigma_0 \langle a^2 \mathcal{V}\rangle  + \langle \mathcal{V}\rangle - \frac{d}{dt}\langle a^2 \mathcal{V}\rangle\bigg]\ .
\end{align}

\section*{References}
\providecommand{\newblock}{}

\end{document}